\documentclass[11pt,a4paper]{article}
\usepackage{jheppub_kim}

\usepackage{multirow, graphicx,amssymb,url,mathrsfs,amsmath}
\usepackage{wrapfig,boxedminipage,setspace,subfigure,epsfig}
\usepackage{amsxtra,amstext,latexsym,dsfont,amsfonts}

\newcommand{\be}{\begin{equation}}
\newcommand{\ee}{\end{equation}}
\newcommand{\bea}{\begin{eqnarray}}
\newcommand{\eea}{\end{eqnarray}}
\newcommand{\ba}{\begin{array}}
\newcommand{\ea}{\end{array}}

\renewcommand{\a}{\alpha}      \renewcommand{\b}{\beta}
\renewcommand{\d}{\delta}      
\renewcommand{\l}{\lambda}

\newcommand{\cala}{\mbox{${\cal A}$}} 
 
 \newcommand{\calf}{\mbox{${\cal F}$}}

 \newcommand{\call}{\mbox{${\cal L}$}}
 
\newcommand{\calo}{\mbox{${\cal O}$}}

\newcommand{\nn}{\nonumber}

\newcommand{\ra}{\rightarrow} 

\newcommand{\lra}{\Longrightarrow}

\newcommand{\dd}{\mathrm{d}}

\newcommand{\tA}{{\tilde{A}}}

\title{The Baryonic Phase in Holographic Descriptions of the QCD Phase Diagram}

\author[a]{Nick Evans,}
\author[a,b]{Keun-Young Kim,}
\author[a]{Maria Magou,}
\author[c]{Yunseok Seo,}
\author[c]{and Sang-Jin Sin}

\emailAdd{evans@soton.ac.kr}
\emailAdd{K.Y.Kim@uva.nl}
\emailAdd{mm21g08@soton.ac.uk}
\emailAdd{yseo@hanyang.ac.kr}
\emailAdd{sjsin@hanyang.ac.kr}

\affiliation[a]{ School of Physics and Astronomy, University of
Southampton, \\ Southampton SO17 1BJ, UK.
}

\affiliation[b]{Institute for Theoretical Physics University of Amsterdam, Science Park 904, \\ Postbus 94485, 1090 GL Amsterdam, The Netherlands}

\affiliation[c]{ Department of Physics, Hanyang University, Seoul 133-791, Korea }

\abstract{
We study holographic models of the QCD temperature-chemical potential phase diagram
based on the D3/D7 system with chiral symmetry breaking. 
The baryonic phase may be included through linked D5-D7 systems.
In a previous analysis of a model with a running gauge coupling a baryonic phase was shown to exist to
arbitrarily large chemical potential. Here we explore this phase in a more generic phenomenological
setting with a step function dilaton profile. The change in dilaton generates a linear confining 
$\bar{q}q$ potential and opposes the screening effect of temperature.
We show that the persistence of the baryonic phase
depends on the step size and that QCD-like phase diagrams can be described.  The 
baryonic phase's existence is qualitatively linked to the existence of confinement in Wilson
loop computations in the background.
  }


\subheader{
           SHEP-12-14
           }

\begin{document}

\maketitle

\section{Introduction}

Holography \cite{Maldacena:1997re,Witten:1998qj} has provided a new and powerful description of strongly coupled gauge theories.
Although a true holographic description of QCD has not been found, many insights have
nevertheless been obtained through understanding confinement \cite{Rey:1998ik,Maldacena:1998im}, chiral symmetry breaking
\cite{Babington:2003vm,Kruczenski:2003uq,Sakai:2004cn},
phenomenological models of mesons such as AdS/QCD \cite{Erlich:2005qh,DaRold:2005zs}, and the transport properties 
of non-abelian plasmas \cite{Policastro:2001yc}. 
For further examples, see \cite{Erdmenger:2007cm, CasalderreySolana:2011us} and references therein. Recently
\cite{Evans:2011eu,Gwak:2012ht} we have been interested in finding a holographic description of the QCD phase diagram and
we will continue those studies here.

The QCD phase diagram is characterized by the quark and baryon densities and the chiral condensate.
It is therefore sensible to begin with a simple holographic model that encodes the physics of
quarks and chiral symmetry breaking. The D3/D7 system is the simplest such example~\cite{Karch:2002sh, Grana:2001xn,Bertolini:2001qa,Kruczenski:2003be, Erdmenger:2007cm}. 
It describes the SU($N_c)$ ${\cal N}=4$
super Yang-Mills theory with $N_f$ quark hypermultiplets. In the quenched approximation the theory is
conformal and on the gravity side is described by probe D7 branes in $AdS_5 \times S^5$. The theory is 3+1 dimensional
at all energy scales and has a conformal UV in which the identification of the operator matching 
between the field theory and the gravity description is clean.

\begin{figure}[]
\centering
  \subfigure[$T$-$\mu$ phase diagram of D3/D7 system with a background magnetic field \cite{Evans:2010iy}. 
    The inset diagrams are representative  probe brane embeddings (dotted lines), 
    where a black disk represents a black hole.]
   {\includegraphics[width=7cm]{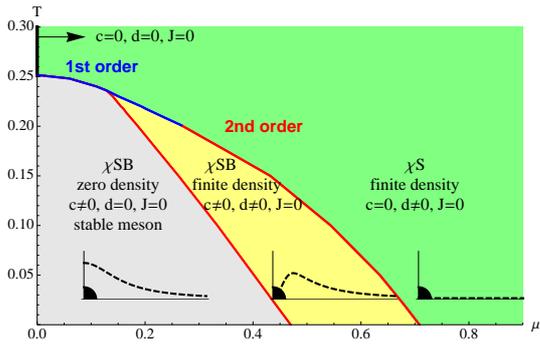} \label{Tvsmu}} \quad
     \subfigure[Phase diagram of a model \cite{Evans:2011eu} with a phenomenological dilaton profile that mimics QCD. ]
   {\includegraphics[width=7cm]{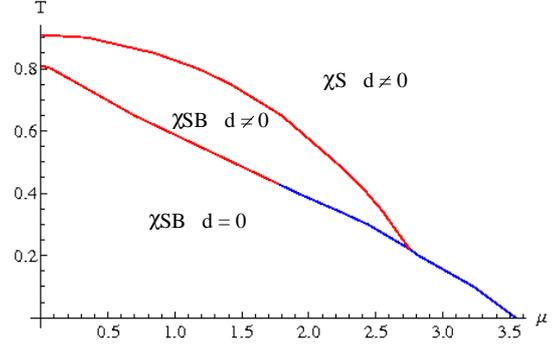} \label{baryone} }
  \subfigure[The phase diagram of the massless axion/dilaton gauge theory in \cite{Gwak:2012ht}
  showing a baryonic phase.]
   {\includegraphics[width=7cm]{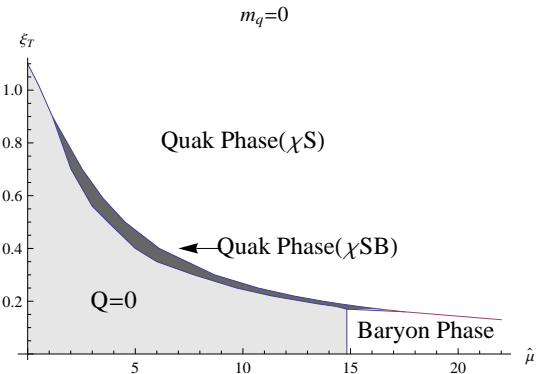} \label{baryone} }
\subfigure[The QCD-like phase diagram for a step function dilaton computed in this paper: $A=10, \l=1.715$]
   {\includegraphics[width=7cm]{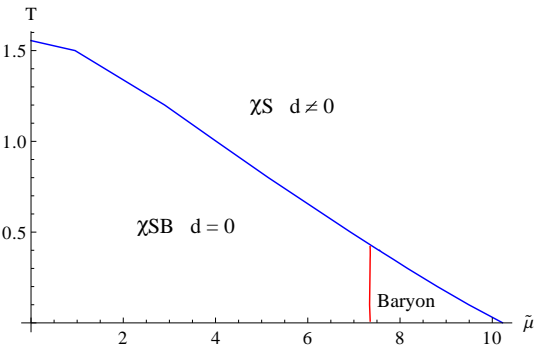} \label{baryone} }   
  \caption{{Example holographic phase diagrams (see the appropriate references for conventions): red lines are 
  first order transitions and blue second order. Phases are labelled by whether chiral symmetry is broken or not
  and whether there is a quark density, $d$, present. In each case the vertical axis is the size of the black hole 
  in the geometry which measures the temperature T.
           }} \label{phase0}
\end{figure}

The simplest example of chiral symmetry
breaking in the D3/D7 system is found by imposing a background magnetic field\cite{Filev:2007gb}. 
In \cite{Evans:2010iy} 
some of the current authors explored the temperature-chemical potential phase diagram of this top down system
which was surprisingly complex. The phase diagram is shown in Fig \ref{phase0}a  
The chiral restoration transition was found to be first order with temperature and second
order with density. A critical point lies between these regimes. In addition there is an extra transition 
associated with the formation of density and the mesons of the system melting into the background plasma \cite{Peeters:2006iu,Hoyos:2006gb}.
In places in phase
space this merges with the chiral transition but in other places it separates and can be either first 
or second order. Such a phase with a quark density but chiral symmetry breaking could potentially exist in 
QCD. On the gravity side the phases are the associated pictograms shown in Fig \ref{phase0}: the D7 can lie
flat, fall on to the black hole which provides temperature, or bend in the space to avoid the origin. 
A number of other top-down analyses of related phase diagrams can be found in
\cite{Jensen:2010vd,Jensen:2010ga,Filev:2010pm,Evans:2010hi,Evans:2010xs,Evans:2011mu,Evans:2011tk,Evans:2011zd,Gwak:2012ht}.

Although the phase diagram of the magnetic field case is interesting it does not overlap with
expectations in QCD where the finite temperature transition is believed to be second order and the
finite density transition first order. In \cite{Evans:2011eu} we introduced a more phenomenological model in which we treated the
running of the gauge coupling/dilaton profile as a free function. This was inspired by 
the fact that the DBI action for the probe with the
magnetic field present is equivalent to the same theory with a particular choice of unbackreacted
dilaton. Using an ansatz which steps between a conformal UV and a conformal IR with a larger coupling value
we could reproduce phase diagrams like that for the magnetic field
case but also a wider set of phase diagrams~\cite{Evans:2011eu}. 
To describe one that looked like the QCD phase
diagram we used a further phenomenological freedom. The D3/D7 system if backreacted 
would display an $SO(4)$$\times$$SO(2)$ symmetry in the directions transverse to the D3 rather than
the SO(6) symmetry of AdS$_5\times S^5$ space. We therefore allowed ourselves to introduce a parameter breaking
the symmetry in that fashion in the emblackening factor of the black hole providing temperature.
With this extra feedom we could produce the phase diagram in Fig \ref{phase0}b which 
maps more closely to that expected in QCD. The high T and low $\mu$ transition can therefore be fitted well.
Here we turn to the phase structure in the low T, high $\mu$ regime.

In \cite{Gwak:2012ht} the phase diagram was studied for the holographic description of a theory with a 
running dilaton. The background has an induced vev for both $Tr F^2$ and $Tr \tilde{F}F$;
in the gravity dual these scalar solutions satisfy the equations of motion whilst leaving
a pure AdS background. Quarks are again introduced using a D7 probe.
The phase diagram is shown in Fig \ref{phase0}c and shows the same three
phases as the cases already discussed. The chiral
transition is first order throughout the plane. The extra key component of the analysis in \cite{Gwak:2012ht} was to note that in the geometry with a running
dilaton a baryonic phase was also present. A baryon vertex is described by a D5 brane wrapped on the 
$S^5$ of the AdS$_5 \times S^5$ space~\cite{Witten:1998xy,Callan:1999zf,Seo:2008qc, Kim:2011gw, Jo:2011em,Seo:2009kg}. In the pure ${\cal N}=4$ theory such vertices shrink to zero size. However in the 
running dilaton geometry the large IR value of the dilaton stabilizes the D5 embedding. Solutions
exist that link the D5 to the D7 brane embedding with a balancing force condition. These configurations
describe the gauge theory with finite baryon density rather than finite quark density. The D7 bends off axis to meet the D5 and so the phase has chiral symmetry broken. This phase
sets in at a particular finite chemical potential and then persists to infinite chemical potential (as
shown in Fig \ref{phase0}c) which is certainly unlike the equivalent phase in QCD (the phase diagram is very similar to that found in the Sakai Sugimoto model in \cite{Bergman:2007wp}) .

The goal of this paper is to combine the approaches of these papers discussed. We will investigate the
baryonic phase in the model with phenomenological freedom in the running coupling. As we had hoped,
we will show that these models can possess a baryonic phase and that for regions of parameter space  
it can be in the finite region one would expect it to be in for QCD.  Our dilaton ansatz
transitions between two conformal regions. In each of those conformal regimes, as in the ${\cal N}=2$ Karch Katz theory,
the baryon vertices are free to shrink. Around the transition radius though there is an extra cost to the D5s
shrinking further since they must encounter the larger dilaton value within. The result is a set of stable D5 configurations 
when the change in dilaton between the two conformal regimes is sufficiently large. These 
configurations are present for low temperatures where the black hole horizon has not eaten the D5s
but not present at large temperatures.
The D7 branes that
introduce the quark fields can end on the D5 baryon vertex when they are available. The D7 branes also prefer not to enter
the interior of the geometry with a large dilaton value. Thus if the dilaton is sufficiently large inside
the D5 radius they will prefer to join the D5. For very large interior dilaton values this occurs for all
values of chemical potential and the baryonic phase exists out to infinite chemical potential as in
the analysis of \cite{Gwak:2012ht} (in that model the dilaton actually diverges in the IR). However we find that
if the change in the dilaton is more modest across the transition then whether the D7 energetically prefer to join
the D5 or end on the black hole horizon must be calculated and depends on the parameters. In particular
we do find ranges of parameter space where the baryonic phase lives in the regime shown in Fig \ref{phase0}d
corresponding to the region expected in QCD. This is the main result of the paper which we will explore in detail.

Finally we briefly explore the link between confinement and the baryonic phase. The quark anti-quark potential
can be computed by drooping a string into the AdS interior \cite{Rey:1998ik,Maldacena:1998im}. The step dilaton profile
induces a linear potential between a quark and an anti-quark when the step size is sufficiently large. The
transition from a Coulomb potential to linear confinement qualitatively happens at the same value of the step size
that induces a baryonic phase in the theory at finite chemical potential as one might expect.

\section{The Holographic Model}

We seek to describe a strongly coupled gauge theory with 
quenched quark fields using the D3/probe-D7 system. A running coupling will be
imposed on the theory through a radial dilaton profile. Finite temperature will be included through an 
AdS-Schwarzchild geometry and baryon number chemical potential via a gauge field in the probe D7 DBI action.
The final ingredient will be to also allow D5 configurations wrapped on the $S^5$ of the geometry
to represent baryon density. We review each of these steps in turn:

\begin{table}[]
\begin{center}
\begin{tabular}{|c|c|c|c|c|c|c|c|c|c|c|}
\hline
&\multicolumn{5}{|c|}{AdS-Black hole} &   \multicolumn{5}{|c|}{$S^5$}\\
\hline
coordinate & $t$ & $x^1$ & $x^2$ & $x^3$  & $ r,w $ &  \multicolumn{5}{|c|}{$S^5$} \\
\hline
Background D3 & $\bullet$ & $\bullet$ & $\bullet$ & $\bullet$  & & \multicolumn{5}{|c|}{} \\
\hline \hline
&\multicolumn{4}{|c|}{}  &\multicolumn{4}{|c|}{$\mathbb{R}^4$} & \multicolumn{2}{|c|}{$\mathbb{R}^2$} \\
\hline
coordinate & $t$ & $x^1$ & $x^2$ & $x^3$  & $\rho$ & \multicolumn{3}{|c|}{$S^3$} &  $L$ & $S^1$ \\
\hline
Flavor D7 &$\bullet$&$\bullet$&$\bullet$&$\bullet$&$\bullet$&$\bullet$&$\bullet$&$\bullet$ & & \\
\hline \hline
&\multicolumn{4}{|c|}{}  & $\mathbb{R}$ &\multicolumn{5}{|c|}{$S^5$} \\
\hline
coordinate & $t$ & $x^1$ & $x^2$ & $x^3$  & $ \xi $ & $\theta$ & \multicolumn{4}{|c|}{$ S^4$} \\
\hline
Baryon vertex D5 & $\bullet$ & & & & & $\bullet$ & $\bullet$ & $\bullet$ & $\bullet$  & $\bullet$ \\
\hline \hline
\end{tabular}
\end{center}\caption{The brane profile showing the coordinates we use: the background D3, the compact D5 and the probe D7 \label{branes}}
\end{table}

\subsection{The Background} 

Our geometry will be the AdS-Schwarzschild black hole metric in Einstein frame which reads
\begin{equation}
\dd s^2 =  \frac{r^2}{R^2}\left( -f \dd t^2
+ \dd \vec{x}_3^3\right) + \frac{R^2}{r^2} \frac{\dd r^2}{f} + R^2 \dd\Omega_5^2  \,, \label{gE}
\end{equation}
where $R$ is the AdS radius ($R^4 = 4\pi g_s N_c \a'^2$) and $f$ is the emblackening factor,
\begin{equation}
 f(r) = 1-\frac{r_H^4}{r^4} \,, \quad \qquad r_H = \pi R^2 T \,,
\end{equation}
with the temperature $T$. 
There is also the self-dual RR five-form $G_{(5)}$ 
\begin{equation} \label{G5}
  G_{(5)} = (1+*) \dd t \wedge \dd \vec{x} \wedge \dd (g_{xx}) \,.
\end{equation}
We will impose an unbackreacted dilaton profile in the AdS-Schwarzschild geometry
to allow us to explore a range of gauge coupling runnings in the gauge theory. In
particular we will use the simple step-function form
\begin{equation} \label{dila1}
  e^{\phi}  = A+1-A\tanh(r-\l) \,,
\end{equation}
where $A$ is the amount of $e^{\phi}$'s change at the position $\l$.

The rescaling of the radius $(r)$ coordinate~\cite{Babington:2003vm}
\begin{equation}
  \frac{dr}{r\sqrt{f}} \equiv \frac{dw}{w} \ \lra \
  w = \frac{1}{\sqrt{2}}\sqrt{r^2 + \sqrt{r^4 - r_H^4}}\ , \quad  
  r = \sqrt{\frac{w^4+w_H^4}{w^2}} 
\end{equation}
with $\sqrt{2}w_H = r_H$, makes the $\mathbb{R}^6$ structure in the transverse space explicit:
\begin{equation}
   \dd s^2 =  \frac{w^2}{R^2}(- g_t \dd t^2 + g_x \dd\vec{x}^2)
     + \frac{R^2}{w^2} (\dd w^2 +  \dd \Omega_5^2 ) \,,
\end{equation}
where
\begin{equation}
\begin{split}
& g_t = \frac{(w^4 - w_H^4)^2}{ w^4 (w^4+w_H^4)}\,,  \quad
g_x  = \frac{w^4 + w_H^4}{ w^4} \,.
\end{split}
\end{equation}

\subsection{D7 flavor brane and quark phase}

We include a small number ($N_f$) of quark flavours by placing probe D7 branes into the background
we have constructed. For this purpose, it is convenient to parameterize $\mathbb{R}^6$ so that $\mathbb{R}^4 \times \mathbb{R}^2$ is explicit (see the middle of Table \ref{branes}):   
\begin{equation} \label{Para1}
  \dd w^2 + w^2 \dd \Omega_5^2 = \dd\rho^2 + \rho^2 \dd\Omega_3^2
         + \dd L^2 + L^2 \dd\Omega_1^2 \,,
\end{equation}
where the D7 brane lies in the ($t,\vec{x},\rho,\Omega_3$) directions
and $L$ and $\rho$ are related to $w,\tilde{\theta}$
\begin{equation} \label{thetatilde}
w = \sqrt{\rho^2 + L^2}\,,  \quad \rho = w \cos\tilde{\theta} \,,
  \quad L = w \sin\tilde{\theta} \,,
\end{equation}
where $0 \le \tilde{\theta} \le \pi/2$ and is different from $0 \le \theta \le \pi$ in the bottom of Table \ref{branes}. 

The action for a D7 brane is given by the DBI action without the Wess-Zumino term
\begin{equation}
 S_{D7}=-T_7\int{\dd^8 \xi e^{\phi} \sqrt{-\text{det}(P[G]_{\alpha\beta}+(2 \pi \alpha')F_{\alpha \b})}}
\end{equation}
where $T_{D7}= \frac{1}{g_{UV}^2 (2\pi)^7 a'^{4}}$ and 
\begin{equation}
e^\phi =  A + 1 - A \tanh\left[
 \left(\sqrt{\frac{(\rho^2 + L^2)^2+w_H^4}{\rho^2 + L^2}} - \lambda\right) \right]\,. \label{beta}
\end{equation}
We add a chemical potential to our system by allowing $\tA_t \equiv 2 \pi A_t(\rho)\neq 0$, and assume that the D7 brane embedding profile is none trivial on $\rho$ only, $L= L(\rho)$.
The action becomes:
\begin{eqnarray}
\begin{split}
 S_{D7}&=\overline{T}_{D7}\int{\dd^4 x \dd\rho \mathcal{L}_{D7}}  \,, \\
 \mathcal{L}_{D7} &= - e^\phi \rho^3 \sqrt{g_t g_x^3 (1+L'^2)-\frac{g_x^3}{e^\phi}\tilde{A}_t^{'2}}
\end{split}  
\end{eqnarray}
where $\overline{T}_{D7}= T_{D7}N_f\Omega_3 $. 

Since $A_t$ is a cyclic coordinate, there is the conserved charge defined by
\begin{equation} \label{DefJ}
  \tilde{d} \equiv \frac{\partial\call_{D7}}{\partial \tA'_t} \,,
\end{equation}
in terms of which, the Wick-rotated (Euclidean, $t \ra -it_E$ ) Legendre transformed action ($\tA_t' \ra \tilde{d}$) is defined by 
\begin{equation} \label{F1}
\begin{split}
  S^{E, LT}_{D7} &= \overline{T}_{D7} \int \dd t_E \dd \vec{x} \int \dd \rho \call^{E, LT}_{D7} \\
  \call^{E,LT}_{D7} &=  \tA'_t \tilde{d} - \call_{D7}  = e^\phi \frac{w^4-w_H^4}{w^4}
         \sqrt{1+L'(\rho)^2}\sqrt{\left(\frac{w^4+w_H^4}{w^4}\right)^2 \rho^6+\frac{w^4}{w^4+w_H^4} \frac{\tilde{d}^2}{e^\phi}}
\end{split}  
\end{equation}
where the following relation obtained in \eqref{DefJ} is used.
\begin{equation} \label{At1}
  \tilde{A}'_t = \tilde{d} \frac{w^4-w_H^4}{w^4+w_H^4}\sqrt{\frac{1+L'^2}{\left(\frac{w^4+w_H^4}{w^4}\right)^2\rho^6+\frac{w^4}{w^4+w_H^4}\frac{\tilde{d}^2}{e^\phi}}} \,.
\end{equation}

We solve the equation of motion for $L(\rho)$ minimizing the action \eqref{F1} for a given $\tilde{d}$. The equation is second order and we impose two boundary conditions: $L(\infty) = m$ and the IR ($\rho \ra \rho_H$) regularity. Asymptotically at large $\rho$ the embedding field $L$ takes the form $L = m + c/ \rho^2+...$ with $m$ and $c$ proportional to the quark mass and condensate bilinear. By plugging the classical solution $L(\rho)$ into \eqref{At1}, we also get the solution $A_t$, for which the large $\rho$ behaviour is $\tA_t = \tilde{\mu} -\frac{\tilde{d}}{2\rho^2}+ ...$ with $\tilde{\mu}$ and $\tilde{d}$ proportional to the chemical potential and quark density. To compute the chemical potential, $\tilde{\mu}$, we simply 
integrate the right hand side of \eqref{At1}, i.e.
\begin{equation} \label{mud}
 \tilde{\mu}=\int_{\rho_H}^{\infty} \dd\rho \, \tilde{d} \, \frac{w^4-w_H^4}{w^4+w_H^4}\sqrt{\frac{1+L'^2}{\left(\frac{w^4+w_H^4}{w^4}\right)^2\rho^6+\frac{w^4}{w^4+w_H^4}\frac{\tilde{d}^2}{e^\phi}}} \,,
\end{equation}
where $\tA_t(\rho_H) = 0$ by a regularity condition at horizon.

The field theory free energy density ($\calf$) is holographically identified as 
\begin{equation}
  \calf = \overline{T}_{D7} \int_{\rho_H}^\infty \dd \rho \left. \call^{E,LT}_{D7} \right|_{\mathrm{on-shell}} \,, 
\end{equation}
where $\call^{E,LT}_{D7}$ is written in \eqref{F1}. 
The field theory grand potential density ($\Omega$) is identified with the Euclidean original action before the Legendre transformation: 
\begin{equation} \label{Grand}
\begin{split}
\Omega &= \overline{T}_{D7} \int_{\rho_H}^\infty \dd \rho \left. \call^{E}_{D7} \right|_{\mathrm{on-shell}} \\
&=\overline{T}_{D7} \int_{\rho_H}^{\infty} 
\dd \rho \rho^6  e^\phi \frac{w^4-w_H^4}{w^4}
\left(\frac{w^4+w_H^4}{w^4}\right)^2
\sqrt{1+L'^2}\sqrt{\frac{1}{\left(\frac{w^4+w_H^4}{w^4}\right)^2\rho^6+\frac{w^4}{w^4+w_H^4}\frac{\tilde{d}^2}{e^\phi}}} \,.
\end{split}
\end{equation}
In both cases, we added the counterterm $\sim - \frac{1}{4\rho^4_{\mathrm{cut-off}}}$ to renormalize the actions.\footnote{Since we consider 
only the massless case, $L(\infty) \ra 0 $, the counter term is simple. 
In general, the counter term is a function of $m$ and $c$.}

\subsection{D5 Baryon Vertex}

The D7 world volume gauge field (dual to the chemical potential) has to be sourced by strings
(quarks in the gauge theory). The string endpoints on the D7 world volume are point charges and the world volume gauge field $A_t$ couples to this point source. The other end of the strings must end on either the black hole horizon or another brane. In the previous section, we considered the first case: the D7 brane touches the black hole horizon and the source for the gauge field
are behind the horizon. Now we turn to the latter case. 

Baryons, bound states of $N_c$ quarks, are described in AdS$_5\times S^5$ by a baryon vertex, a D5 brane wrapped on the five sphere with $N_c$ fundamental strings attached to it. Therefore, it is natural to consider a baryon vertex as a source of strings coupling to the $A_t$ world volume field of the D7 brane. 

For a D5 brane baryon vertex configuration, it is more useful to parameterize $\mathbb{R}^6$ as 
\begin{equation} \label{Para2}
  \dd w^2 + w^2 \dd \Omega_5^2 = \dd \xi^2 + \xi^2 \left(\dd\theta^2+ \sin{\theta}^2 \dd\Omega_4^2\right) \,,
\end{equation}
rather than \eqref{Para1},  since  the D5 brane will lie in the $(t, \Omega_4)$ directions with a non-trivial profile $\xi(\theta)$. See the bottom of Table \ref{branes}. We renamed $w$ to $\xi$, to 
make clear that the radial coordinate is a function of $\theta$ here, which is different from $w(\rho) = \sqrt{L(\rho)^2 + \rho^2}$ in the D7 brane case. Furthermore, $\theta$ here is also different from $\tilde{\theta}$ in the D7 brane case as noted below \eqref{thetatilde}. The plots in the following section show the
D7 and D5 brane embeddings simultaneously in one plot, which is not, strictly speaking, correct because of these different coordinate systems. The superposition of the two pictures, which only match where the two branes join, is though
helpful to understand the solutions.  

The action for the D5 baryon vertex in the string frame is given by the DBI and the Wess-Zumino term:
\begin{equation}
 S_{D5}=-T_{D5} \int{\dd^6 \xi e^{-\phi} 
 \sqrt{-\text{det}(P[G]_{\alpha\beta}+(2 \pi \alpha')F_{\alpha b})}}+ T_{D5}\int{\dd^6 \xi \cala_{(1)}\wedge G_{(5)}}
\end{equation}
where $T_{D5}= \frac{1}{g_{UV}^2 (2\pi)^5 a'^{3}}$ and $\cala_{(1)}$ is the world volume gauge field one-form on 
the D5 brane, which is different from $A_t$ introduced on the D7 brane. In \eqref{Para2}  the dilaton reads
\begin{equation}
e^\phi =  A + 1 - A \tanh\left[
\left(\sqrt{\frac{ \xi^4+w_H^4}{ \xi^2 }} - \lambda\right) \right]\,. \label{beta}
\end{equation}
Note that the Wess-Zumino term contributes to the action as well due to the coupling of the worldvolume gauge field $\cala_{(1)}$ with the background five-form $G_{(5)}$ \eqref{G5}. A nontrivial temporal gauge field $\cala_t(\theta)$ couples to $N_c$ charge on the D5 brane. With assumptions $\xi = \xi(\theta)$ and $\cala_t = \cala_t(\theta)$, the action is given by:
\begin{eqnarray}
 \begin{split}
 S_{D5} 
 &=\overline{T}_{D5}\int{\dd t \dd\theta \mathcal{L}}_{D 5} \nn \\
 \call_{D5}&= - \sin{\theta}^4 \left(\sqrt{e^\phi} \sqrt{\frac{(\xi^4-w_H^4)^2}{\xi^4(\xi^4+w_H^4)} (\xi^2+\xi'^2)-\frac{1}{e^{\phi}}\tilde{\cala'}_t^2}-4 \tilde{\cala}_t\right) 
 \nonumber \,,
 \end{split}
\end{eqnarray}
where $\overline{T}_{D5}= R^4 \Omega_4 T_{D5}$. Note that we extract $R^4$ to make $\call_{D5}$ dimensionless. 

The equation of motion for the gauge field $\cala_t(\theta)$ takes the form:
\begin{equation} \label{eq:Dtheta}
 \partial_\theta \tilde{D}(\theta) = 4 \sin{\theta}^2 \,,
\end{equation}
where the conjugate momentum $\tilde{D}$ of $\tilde{\cala_t}$ is defined by
\begin{equation} \label{eq:Dtheta1}
 \tilde{D}(\theta) = \frac{\delta \mathcal{L}_{D 5}}{ \delta \tilde{\cala_t}'}  =\frac{\tilde{\cala_t}'\sin^4{\theta}}{\sqrt{e^{\phi}}\sqrt{\frac{(\xi^4-w_H^4)^2}{\xi^4(\xi^4+w_H^4)}(\xi^2+\xi'^2)-\frac{\tilde{\cala'_t}^2}{e^{\phi}}}}
\end{equation}
The general solution of \eqref{eq:Dtheta} is:
\begin{equation} \label{tildeD}
 \tilde{D}(\theta)=\frac{3}{2}(\nu \pi -\theta+\frac{3}{2}\sin{\theta}\cos{\theta}+\sin{\theta}^3\cos{\theta}) \,,
\end{equation}
where $0\leq \nu \leq 1$ is the integration constant and it is related to the number of fundamental strings attached to each pole~\cite{Callan:1999zf}. For our purposes we choose $\nu=0$ which is the case that all the fundamental strings emerge from only one pole of the baryon vertex which we choose to be $\theta=\pi$.
Using \eqref{eq:Dtheta1} we can rewrite our Euclidean baryon vertex Lagrangian as:
\begin{eqnarray}
 \call_{D5}^E =   \sqrt{e^\phi} 
\sqrt{\frac{(\xi^4-w_H^4)^2}{\xi^4(\xi^4+w_H^4)}(\xi^2+\xi'^2)}\sqrt{\tilde{D}(\theta)^2+\sin^8{\theta}} \,,
\end{eqnarray}
where $\tilde{D}(\theta)$ is defined in \eqref{tildeD} with $\nu = 0$. 
This is for one Bayron vertex. The holographic energy density of many non-interacting free baryon vertex system at finite density $n_B$ is 
\begin{equation}
  n_B \overline{T}_{D5} \int \dd \theta \call_{D5}^E \,
\end{equation}
where  $n_B = \frac{n_q}{N_c}$ and $n_q$ is the quark density.

\subsection{Baryon phase: D7 + D5 branes}

The baryon phase is constructed by connecting a D7 flavour brane and D5 baryon vertices by strings between them. 
It can be shown that the strings' tension is 
so strong that they tend to shrink to a point \cite{Gwak:2012ht}, which makes the D7 flavor brane and D5 baryon vertices meet at a point. Therefore, ignoring strings, we start with the D7 - $n_B$ D5 combined system. Its free energy density is 
\begin{align}
 \calf_{B}&=\overline{T}_{D7}\int_{0}^{\infty}{ \dd\rho \mathcal{{L}}_{D7}^{E,LT}}+ \frac{n_q}{N_c} \overline{T}_{D5} \int_{0}^{\pi}{ \dd \theta \mathcal{{L}}}_{D 5}^{E}\\
            &=\overline{T}_{D7}\left(\int_{0}^{\infty}{ \dd\rho \mathcal{{L}}_{D7}^{E,LT}}+\frac{2}{3 \pi } \tilde{d}  \int_{0}^{\pi}{ \dd \theta \mathcal{{L}}}_{D 5}^{E}\right) \,, \label{exd}\\
            &\equiv \overline{T}_{D7} (\tilde{\calf}_{D7} + \tilde{\calf}_{D5} ) \,,
\end{align}
where $\frac{n_q}{2\pi\a' \overline{T}_{D7}} = \tilde{d}$ since $n_q$ is identified with $\frac{1}{V_3}\frac{\d S_{D7}}{\d A_t(\infty)} $ ($V_3$ is the three dimensional volume).
Let us consider, at a fixed finite density $\tilde{d}$, the configuration of the D7 brane with a fixed boundary value $L(\infty) = m$ and the D5 brane with a fixed $\xi(0) = \xi_0$.  The two brane embeddings have to meet at $\rho=0$ and $\theta = \pi$, i.e. $L(0) = \xi(\pi) \equiv w_0$. There are infinitely many configurations satisfying this condition, parameterized by $w_0$. To find out the lowest energy configuration, we vary the total free energy.  
\begin{equation}
\begin{split}
  \d \calf_{B} & \sim \int_{0}^{\infty}{ \dd\rho (\mathrm{EOM_L} ) \d L }
  + \left. \frac{\partial \call_{D7}^{E,LT} }{\partial L'} \d L \right|^{\infty}_{0} \\
  & \quad
   +\frac{2}{3 \pi } \tilde{d}  \int_{0}^{\pi} (\mathrm{EOM_\xi} ) \d \xi 
   +\left. \frac{2}{3 \pi } \tilde{d}  \frac{\partial \call_{D5}^{E} }{\partial \xi'} \d \xi \right|^{\pi}_{0} \\
  & \sim  -\left. \frac{\partial \call_{D7}^{E,LT} }{\partial L'} \d L \right|_{\rho=0} +\left. \frac{2}{3 \pi } \tilde{d}  \frac{\partial \call_{D5}^{E} }{\partial \xi'} \d \xi \right|_{\theta=\pi} \,,
\end{split}
\end{equation}
where the EOM$_L$ and EOM$_\xi$ are the equations of motion of $L$ and $\xi$ respectively. They vanish since we consider only the solutions of the equation piecewise. 
$\d L(\infty) = \d \xi(0) = 0$ by our boundary condition.
At the matching point, $\d L = \d \xi$, so the condition is reduced to 
\begin{equation}
    \left. \frac{\partial \call_{D7}^{E,LT} }{\partial L'} \right|_{\rho=0} = \left. \frac{2}{3 \pi } \tilde{d}  \frac{\partial \call_{D5}^{E} }{\partial \xi'}  \right|_{\theta=\pi} \quad \Rightarrow \quad  L'(0) = \frac{\xi'(\pi)}{\xi(\pi)} \,,
\end{equation}
which is called a force balancing condition~\cite{Seo:2009kg,Gwak:2012ht}.

In the baryon phase, the chemical potential has an extra contribution 
from D5 branes. 
\begin{equation}
  \tilde{\mu} \equiv \frac{1}{\overline{T}_{D7}} \frac{\partial \calf_B}{\partial \tilde{d}}  = \tilde{\mu}_{D7} + \frac{2}{3\pi}\int_{0}^{\pi}{ \dd \theta \mathcal{{L}}}_{D 5}^{E} \,. \label{muB1}
\end{equation}
where $\tilde{\mu}_{D7}$ means \eqref{mud} now integrated from the D5/D7 join to infinity. Note the extra term from the D5 originates from \eqref{exd} where there is an extra coefficient of $\tilde{d}$.

However, both in the quark phase and the baryon phase, the grand potential is computed as a D7 brane Euclidean on-shell action, which is written explicitly in \eqref{Grand}. The D5 brane action does not explicitly contribute in the baryon phase  because 
\begin{equation}
\begin{split}
  \tilde{\Omega}_B \equiv \frac{\Omega_B}{\overline{T}_{D7}}&= \tilde{\calf}_B - \tilde{\mu} \tilde{d} \\ 
  & = \tilde{\calf}_{D7} + \tilde{\calf}_{D5} - \left(\tilde{\mu}_{D7} + \frac{2}{3\pi }\int_{0}^{\pi}{ \dd \theta \mathcal{{L}}}_{D 5}^{E}\right)\tilde{d} \\
  &= \tilde{\calf}_{D7} - \tilde{\mu}_{D7}\tilde{d} \,.
\end{split}
\end{equation}
However, the D5 brane still implicitly contributes by changing the classical embedding solution. In the following section, we will drop
the subscripts $B$ and $D7$. \\

\section{Phase diagram in grand canonical ensemble}

We will explore the phase diagram of the model with massless quarks.
The phase diagram, neglecting the baryon vertex, was explored in \cite{Evans:2011eu}.
The scale $\lambda$ is the only conformal symmetry breaking scale in the model and so its value can
be scaled (for numerical work we take $\lambda=1.715$ to match previous work). The phase structure
depends on the value of the parameter $A$ that determines how much the dilaton changes between the UV 
and the IR.

 \begin{figure}[]
\centering
  {\includegraphics[width=6cm]{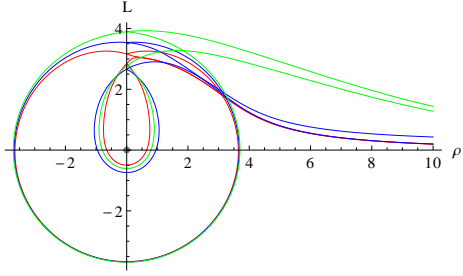} \label{BVEmb}}\quad
  \caption{Examples of the baryon vertex solutions and their behaviour as we increase density - for these 
  parameters there are two vertices for each value of density. 
   The energetically prefered one is always that with the largest radius. The red lines corresponds to $d=0.01$, blue $d=5$ and green $d=1000$. Parameters $w_H=0.1,A=10,\lambda=1.715$.}
            \label{BV1}
\end{figure}
\begin{figure}[]
\centering
  \subfigure[$w_H=0.1$ ]
   {\includegraphics[width=6cm]{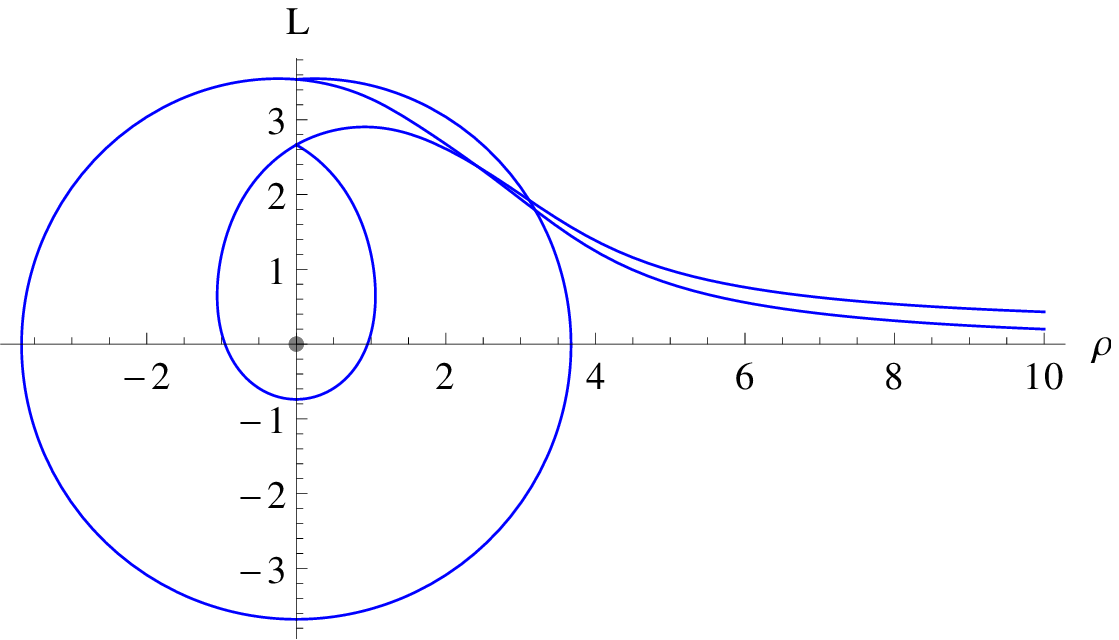} \label{Emb1}} \quad
     \subfigure[$w_H=0.345$ ]
   {\includegraphics[width=6cm]{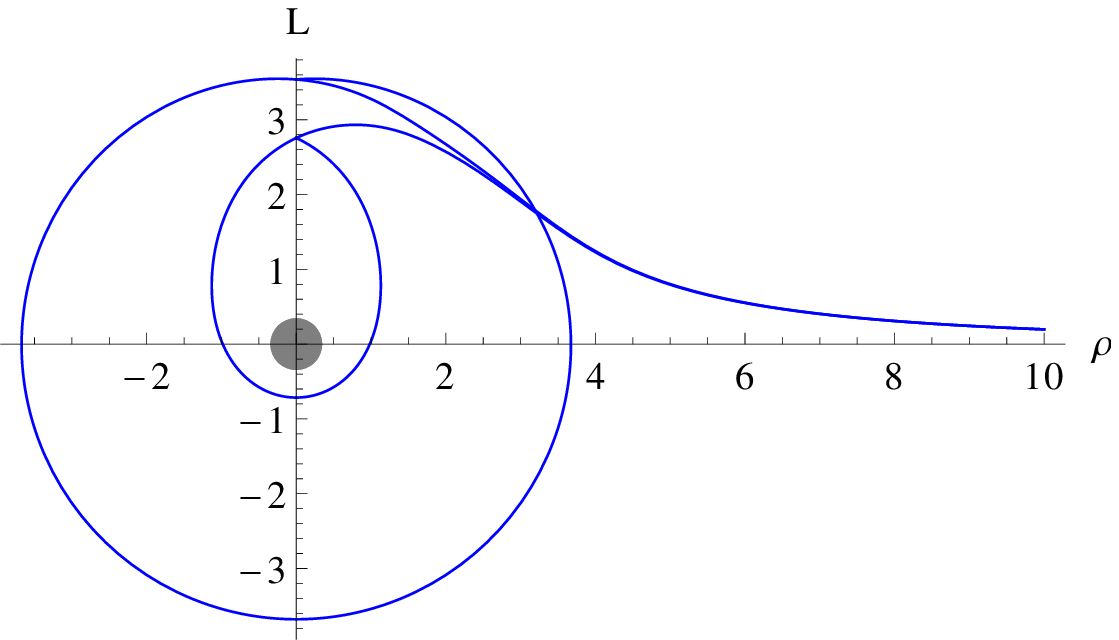} \label{Emb2} } \quad
\subfigure[$w_H=0.8$  ]
   {\includegraphics[width=6cm]{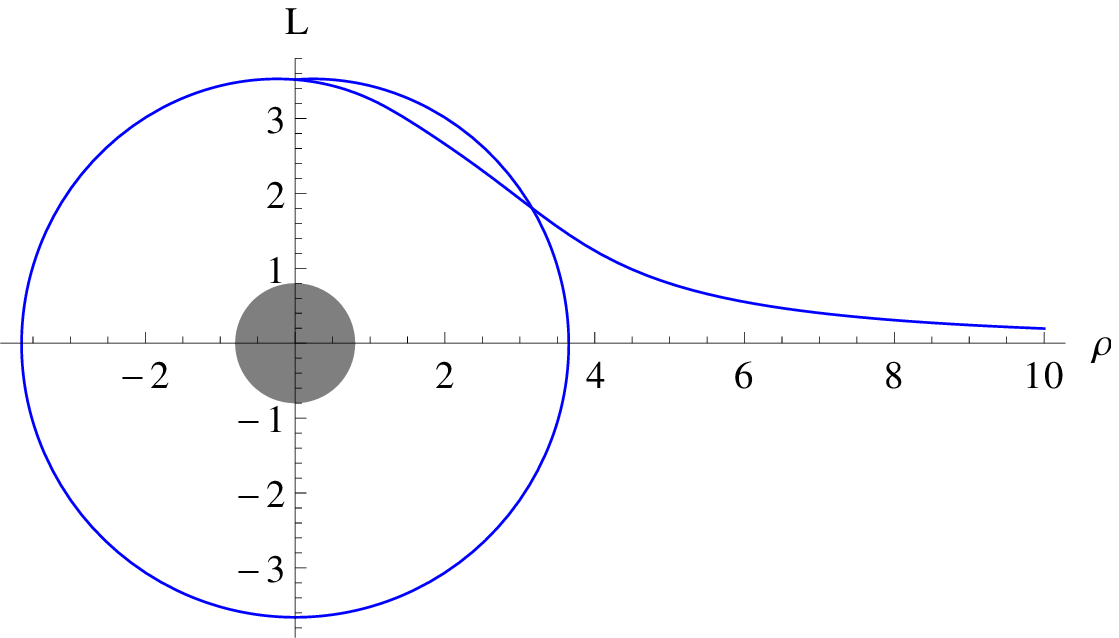} \label{Emb3} } \quad
\subfigure[$w_H=1.5$ ]
   {\includegraphics[width=6cm]{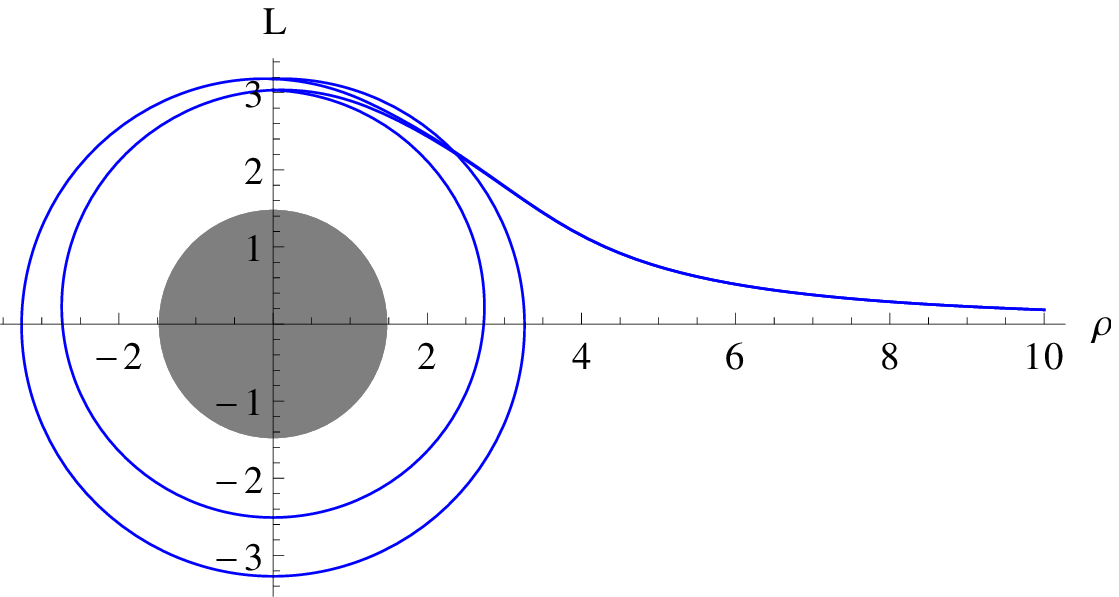} \label{Emb4} }
  \caption{{Behavour of the baryon vertex with temperature for fixed density. There is always one baryon vertex 
  for the temperatures considered and typically two
   as shown. The energetically prefered one is always the one with the largest radius. Parameters $d=5, A=10,  \lambda=1.715$. }
           } \label{BV2}
\end{figure}

For each point in the phase diagram one numerically seeks all possible D7 embedding solutions that asymptote to
$L=0$ at large $\rho$. There is always the flat solution $L(\rho)=0$. There can also be ``Minkowski'' solutions
that end at $\rho=0$ with $L'(0)=0$. Finally there can be embeddings that end on the black hole horizon.
It is convenient to fix the density $d$ and then determine $\mu$ from (\ref{mud}).
The grand potentials  
of the solutions are then compared to determine the prefered phase. 

\begin{figure}[]
\centering
  \subfigure[$w_H=0.1$ ]
   {\includegraphics[width=6cm]{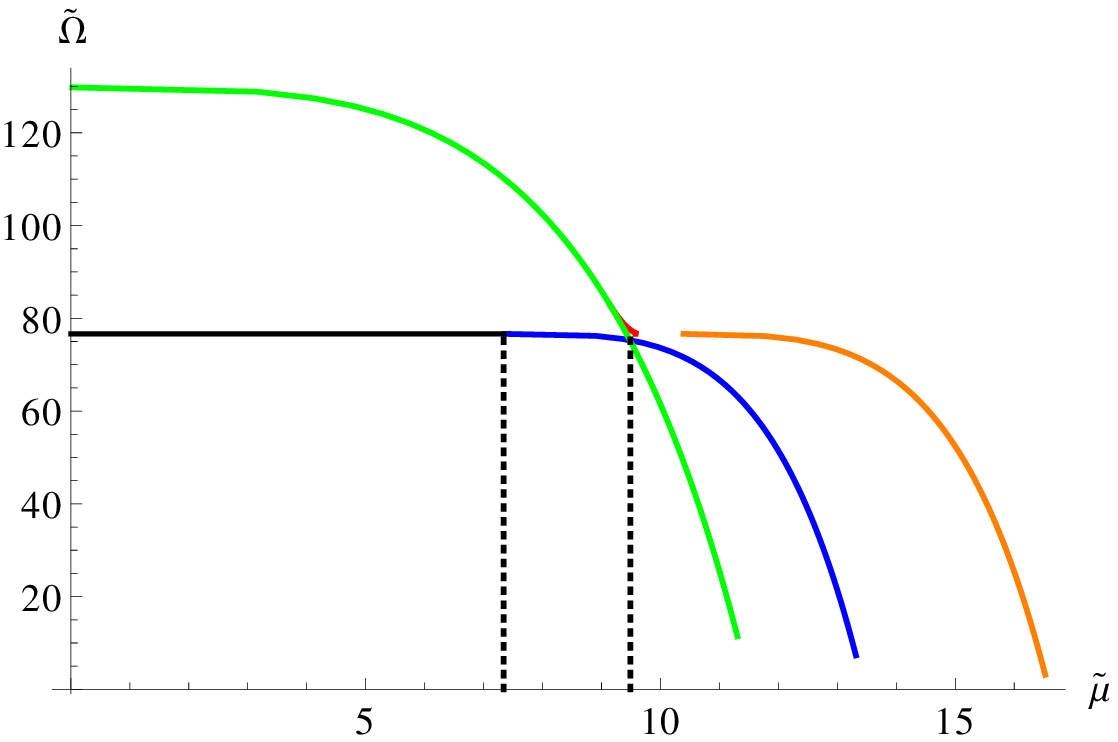} }
     \subfigure[$w_H=0.8$ ]
   {\includegraphics[width=6cm]{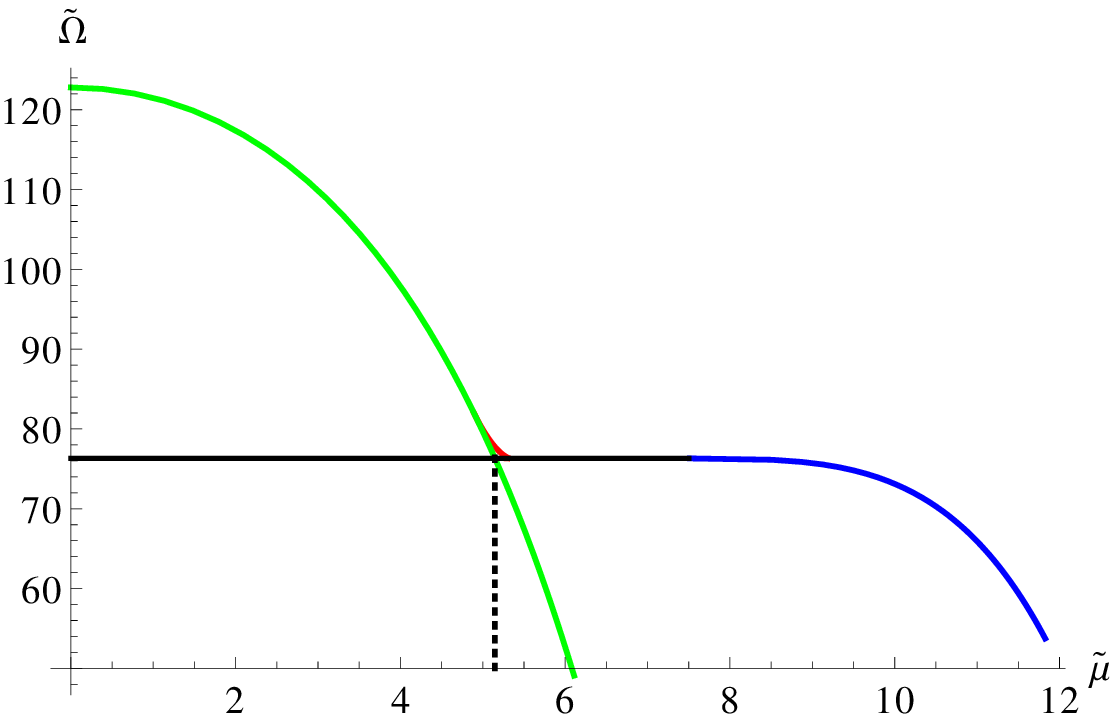} }
    \subfigure[$w_H=0.1$ ]
   {\includegraphics[width=6cm]{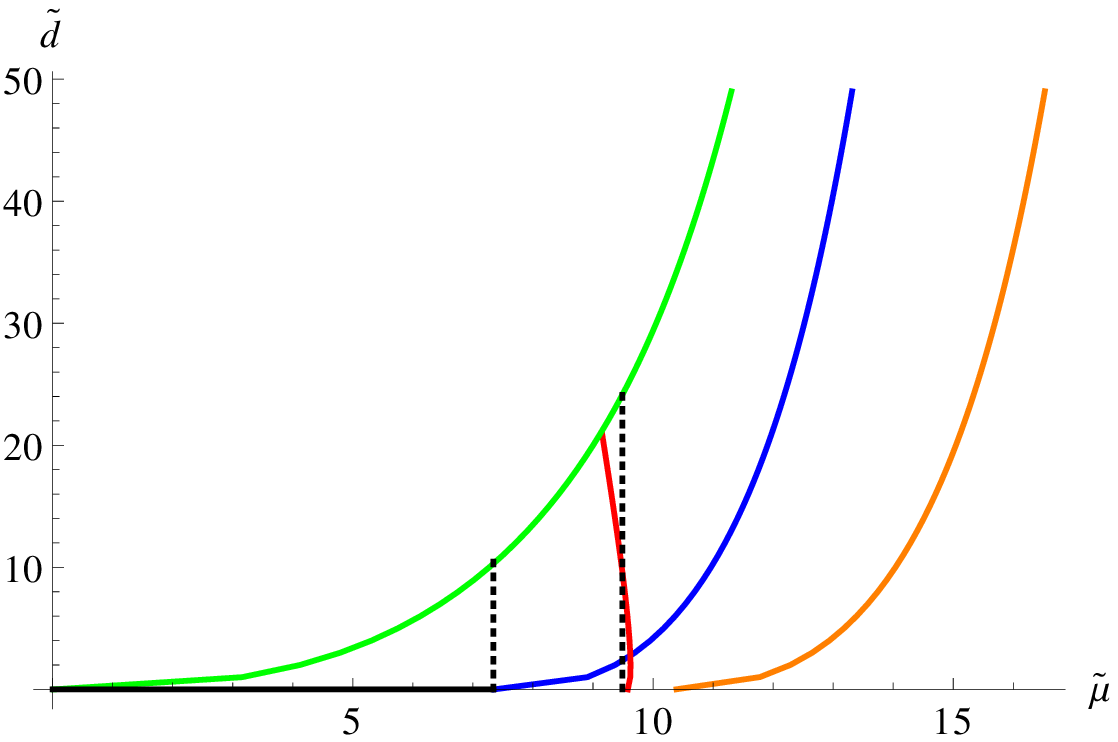} } 
\subfigure[$w_H=0.8$ ]
   {\includegraphics[width=6cm]{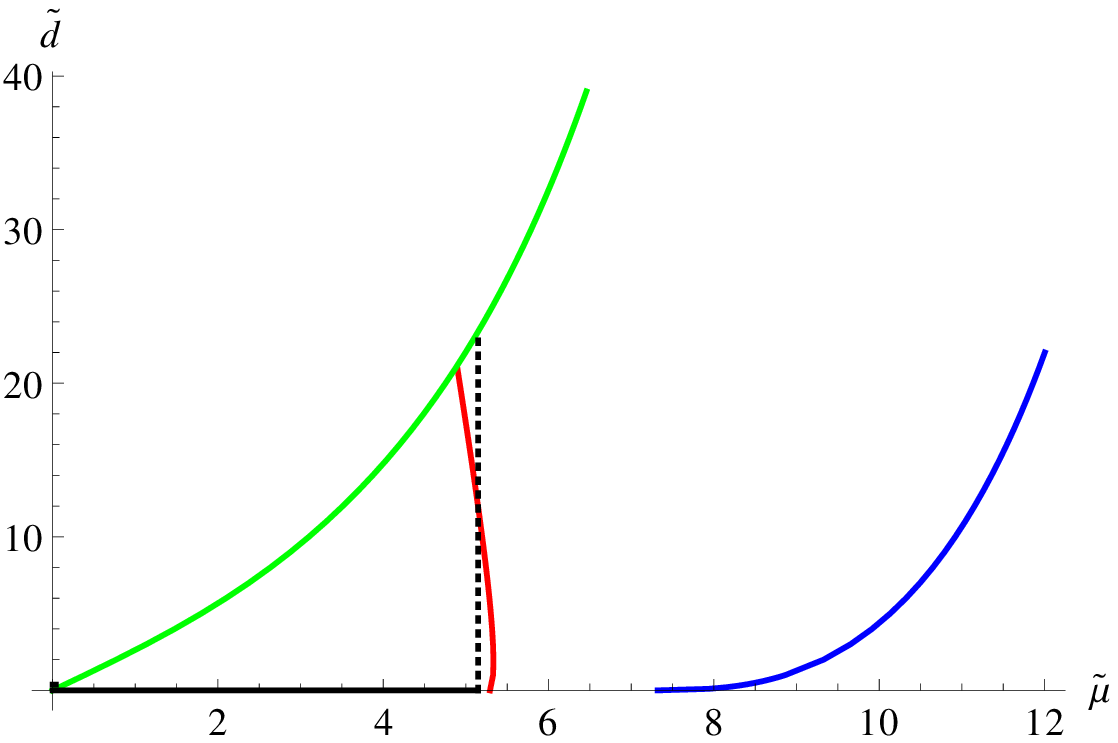} }
   \subfigure[$w_H=0.1$  ]
   {\includegraphics[width=6cm]{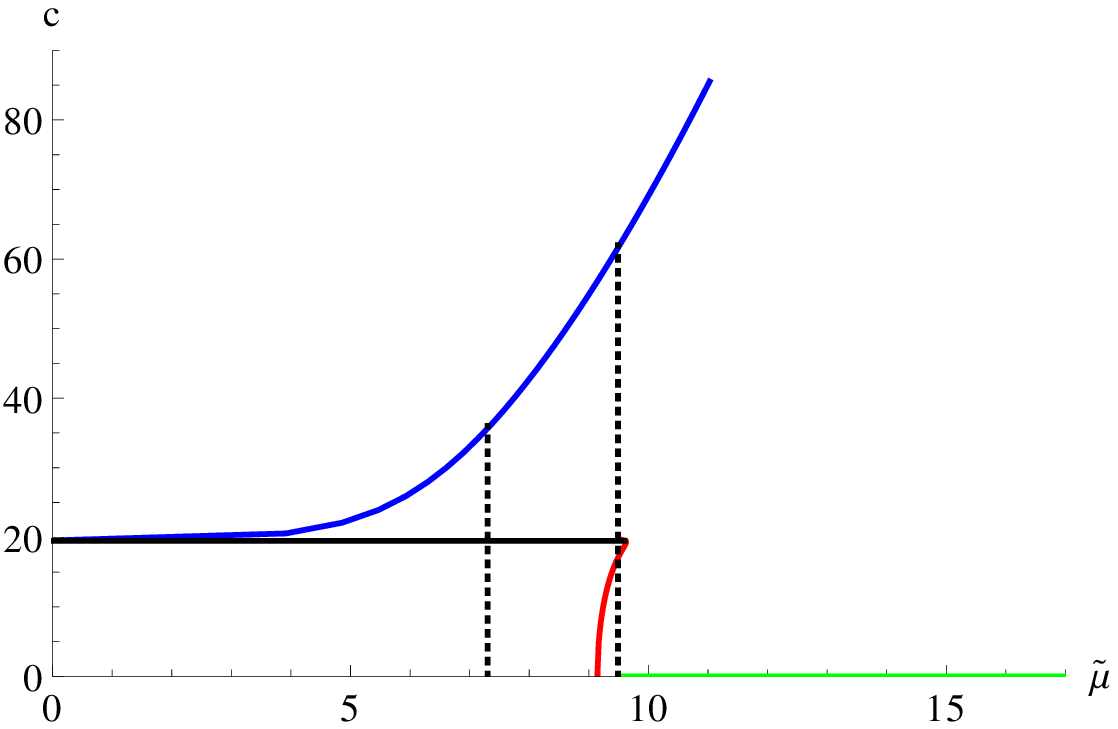} \label{Emb3} } \quad
\subfigure[$w_H=0.8$ ]
   {\includegraphics[width=6cm]{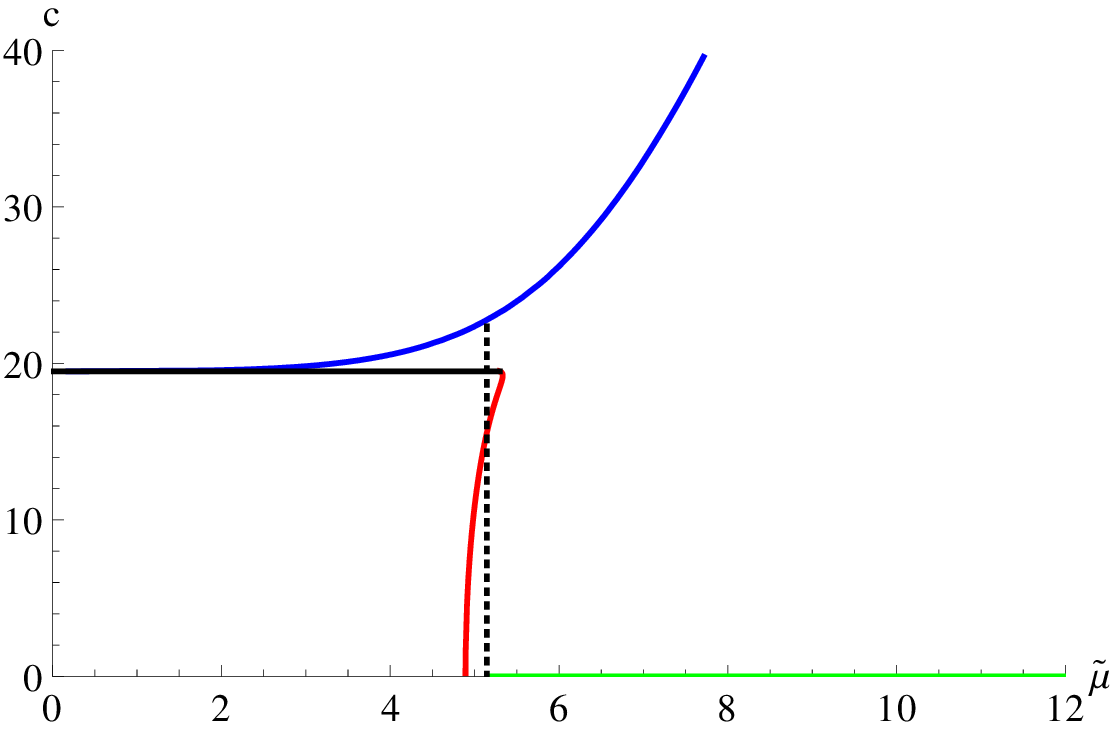} \label{Emb4} }
  \caption{Plots showing the variation of a number of order parameters through the transition regions. The colour
  coding is - Green:flat embedding, Black: Minkowski embedding, Blue: large radius baryon vertex, Orange: small
  radius baryon vertex, Red: black hole embedding. Parameters: $A=10,\lambda=1.715, \Gamma=1$. Top: the grand canonical potential vs chemical potential,
  Middle: $d$-$\mu$ plot, Bottom: condensate $c$ vs  $\mu$.  
     } \label{grand}
\end{figure}

For $A<2.1$ the model is always in the chirally symmetric phase. The massless D7 embedding is 
$L(\rho)=0$ at all temperatures and density. To induce chiral symmetry breaking the cost of
entering the interior volume with the enlarged dilaton must outweigh the cost of bending off axis.
For $A> 2.1$ such chiral symmetry breaking is preferred at low $T$ and $\mu$. In the 
parameter regime $2.1< A < 15$ 
the phase diagram divides into a chiral symmetric phase at high $T, \mu$ and a region with
chiral symmetry breaking at low $T,\mu$. The transition between is first order throughout the phase
diagram. 
For $A>15$ D7 embeddings that end on the black hole play a role in the phase structure 
at high $\mu$ and the phase diagram mutates to the form shown in Fig \ref{phase0}a \cite{Evans:2011eu}. 
The extra phase
has non-zero quark number and chiral symmetry breaking. The transitions between the phases at
high density become second order (the transition with T at $\mu=0$ remains first order no matter how
large $A$).

\begin{figure}[]
\centering
     \subfigure[$A=3$ ]
   {\includegraphics[width=6cm]{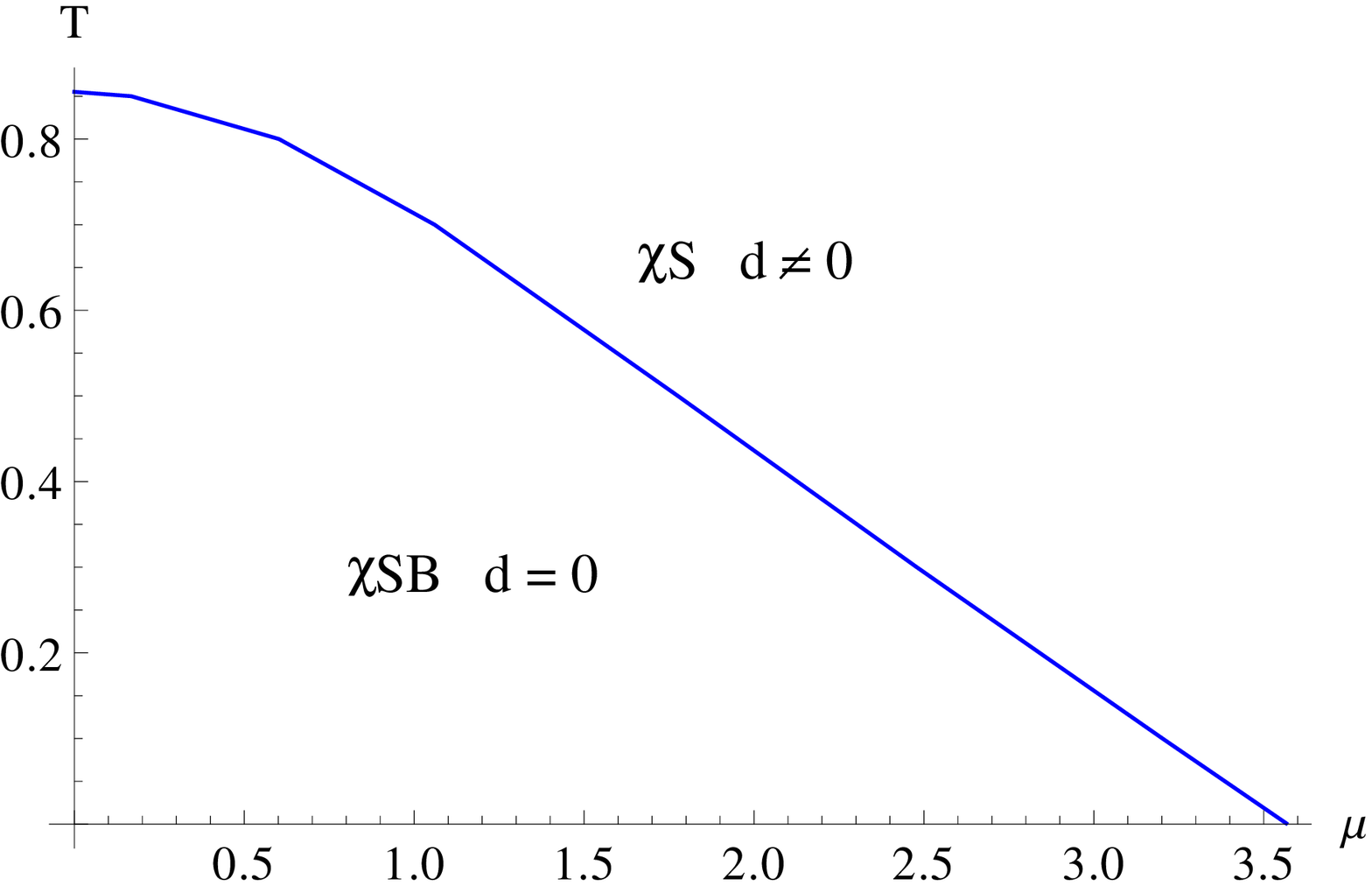} }
    \subfigure[$A=10$ ]
   {\includegraphics[width=6cm]{fig1d.eps} } 
\subfigure[$A=20$ ]
   {\includegraphics[width=6cm]{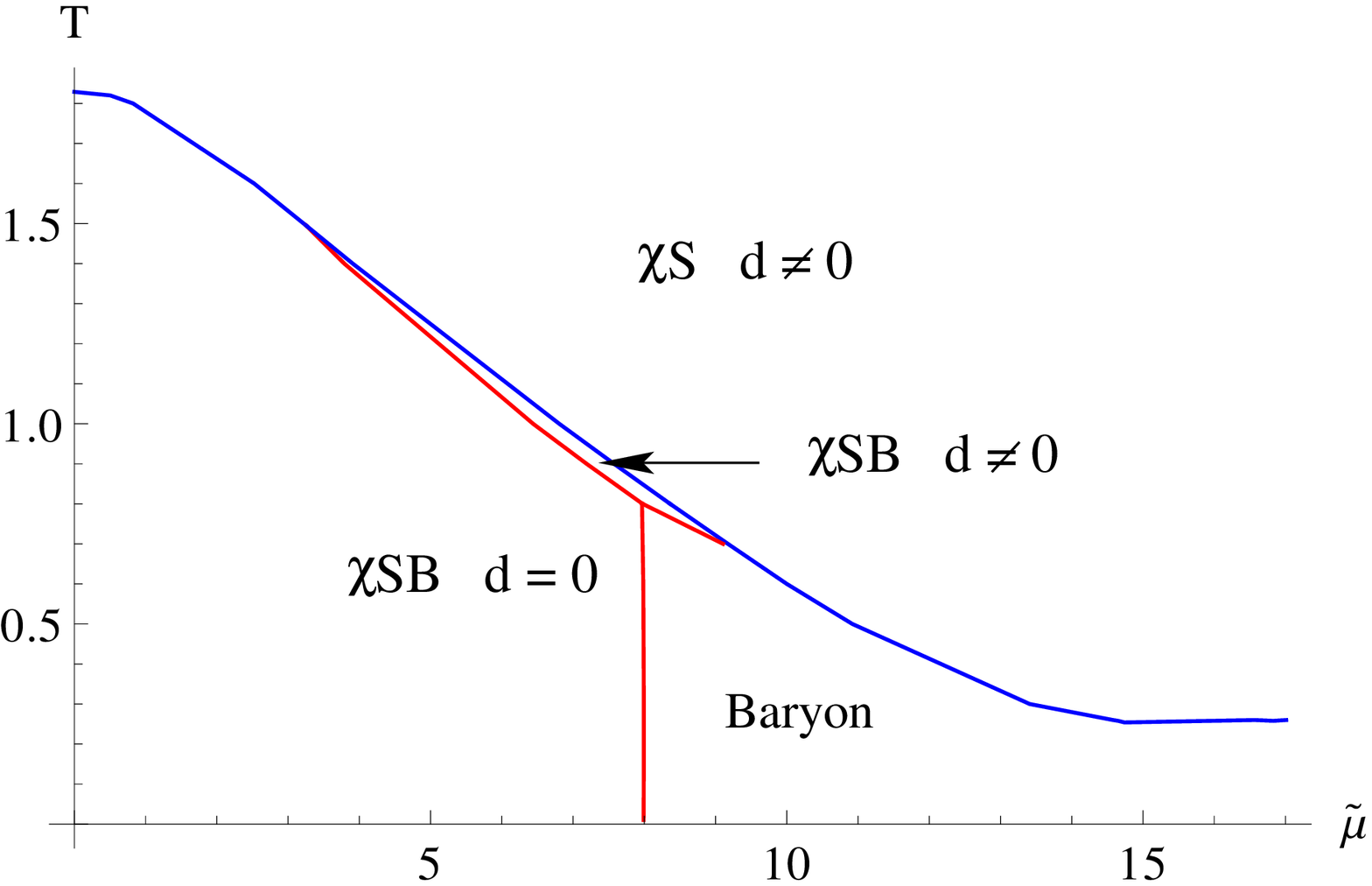} }
  \caption{Phase diagrams as a function of the parameter $A$ in the dilaton profile. The temperature $T$ is in
  units of $w_H$. 
  Transitions marked in blue are first order, those in red second order. The phases are labelled by 
  whether chiral symmetry is broken and whether there is a baryon or quark density.} \label{final}
\end{figure}

In addition to these embeddings we can now seek linked D5-D7 configurations as well with
the same massless asymptotic boundary condition on the D7. Some examples
of these configurations are shown in Fig \ref{BV1} and Fig \ref{BV2}. The D5s are all stabilized for radii
of order $\lambda$. 

Generally the number of baryon vertex solutions 
(linked D5/D7 solutions with massless boundary conditions on the D7) vary with the parameters $A,d$ and $T$. 
For some parameter sets we don't have any baryon vertex solutions (for small $A$'s and densities). 
In Fig \ref{BV1} we show an 
example of a parameter set where we have two baryon vertex solutions. Regardless of
the number of baryon vertex solutions, we have always found
that the energetically favoured solution is the one with the largest radius at the south pole. 
As we increase $A$ the number of solutions increases and the D5 embeddings grow in radius. 
The baryon vertex solutions also change with density. Starting from very small density and increasing 
density the baryon vertex increases its radius rapidly in the beginning before asymptoting to a  slower
growth for larger densities.
At fixed density, increasing temperature typically generates a smaller baryon vertex due to the black hole 
attraction (Fig \ref{BV2}).
 
Our job now is to again compare the Grand Potential energies of all these configurations. Some
examples of that process are shown in Fig \ref{grand}(a,b). The points where transitions occur 
can be read from the energies and then checked against transitions in order parameters of the model.
For example in Fig \ref{grand}(c,d) we plot the quark density against the chemical potential where the
transitions and their orders can also be seen. The parameter $c$ determining the quark condensate
is also shown in Fig \ref{grand}(e,f).

After this work one can construct the full phase diagram which is shown in 
Fig \ref{final} for three choices of the parameter $A$. For small $A$ (but still above the threshold 
for there to be chiral symmetry breaking) the baryonic phase plays no role. The cost of entering the high $A$ area 
is not large enough to discourage the flat embedding. 

At intermediate $A$ we see the baryonic phase enters 
at intermediate chemical potential and low $T$. Fig \ref{final}b looks similar to expectations in QCD
for the position of the baryonic phase and this is the most significant result we present. Note that
the transition to the baryonic phase from the vacuum phase is second order. In QCD it is expected
to be first order at low temperature due to the interactions between the nucleons. At low $\mu$
the phase in QCD can be thought of as sparse liquid droplets of nucleons (ie nuclei) in the 
vacuum. When $\mu$ is sufficient to fill space with nucleons the droplets rearrange themselves into
the hadron gas. The entropy change associated with the internuclear interactions in this rearrangement
generates the first order behaviour. In our analysis
such interactions are neglected so not surprisingly a simple second order transition results. 
Recent work on computing inter-baryon forces using holography are summarized in \cite{Aoki:2012th}.

Finally for large $A$ the baryonic phase is stable out to infinite
chemical potential - the cost of entering the large $A$ region is so great that the flat embedding is always
less prefered than a D5 ending embedding. This matches the model of \cite{Gwak:2012ht} in
which the dilaton diverges at small radius (and also the Sakai Sugimoto phase diagram in \cite{Bergman:2007wp}).
The $A=20$ diagram also displays a small phase region in which black hole embeddings are dominant
corresponding to a chiral symmetry broken phase but with quark density.

\section{The canonical ensemble}

\begin{figure}[] \centering
   {\includegraphics[width=8cm]{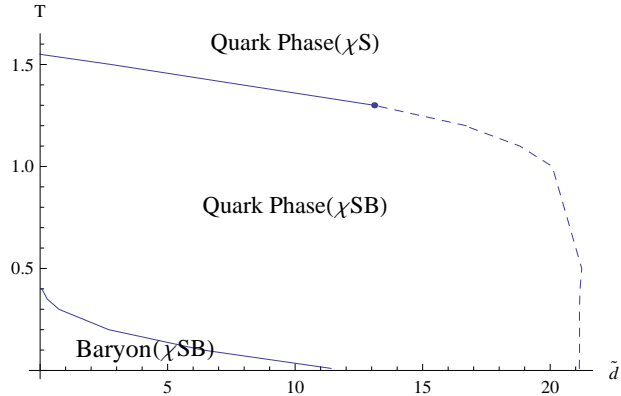} } 
  \caption{The canonical ensemble phase diagram for the case $A=10$. The temperature $T$ is in
  units of $w_H$. Transitions marked in blue are first order, those in red second order. The phases are labelled by 
  whether chiral symmetry is broken and whether there is a baryon or quark density.} \label{cano}
\end{figure}

 The phase diagram of our theory shows some additional structure in the canonical ensemble
that is worth mentioning. We will work just in the case $A=10$. Here we fix the density $d$
rather than the chemical potential $\mu$ and we minimize the free energy rather than the grand potential.
We display the phase diagram in Fig \ref{cano}. The chiral symmetry breaking phase with $d=0$ 
in the grand canonical ensemble lies entirely on the $T$ axis in this plot. Interestingly though
the canonical ensemble phase diagram includes a region with black hole embeddings (the chiral
symmetry breaking, $d \neq0$ phase) which is absent in the grand canonical ensemble. This is
not a discrepancy. At the first order transitions in the grand canonical ensemble there is a jump
in all order parameters including the quark density $d$ - there must therefore be regions of 
the canonical ensemble phase diagram that are not present in the grand canonical ensemble and
indeed are energetically unstable. On any given fixed $T$ line the first order transition from either the
Minkowski embedding or the baryon phase to the symmetric phase ``leaps'' over the black hole embedding
phase of the canonical ensemble. As we saw in Fig \ref{grand}(c) for $A=20$ this third phase does
play a small role in the phase diagram in the grand canoncial ensemble.

\section{Quark-antiquark potential}

We found above that a larger dilaton step-size,$A$, makes a baryonic phase more stable, which implies that a large $A$ triggers a confinement transition as well as chiral symmetry breaking. Therefore, it is interesting to 
study the confinement/deconfinement transition property by other methods.
One of the standard tools is the Wilson-Polyakov loop, which can be obtained holographically by computing the on-shell Nambu-Goto action\cite{Rey:1998ik,Maldacena:1998im,Rey:1998bq,Brandhuber:1998bs,Ghoroku:2005tf}. 

\begin{equation} \label{NG1}
  S_{NG} = \frac{1}{2\pi\a'} \int \dd r e^{\phi/2} \sqrt{1
  + \left( \frac{r}{R} \right)^4 f(r) x'^2},
\end{equation}
where the static string worldsheet is parametrized as $x(r)$ and $x$ is 
one of the field theory spacial direction. There are two types of solutions: a pair of parallel strings  and a U-shaped string. In principle, the string end points can end on the deep IR bottom of the D7 probe brane. Such a non-trivial embedding is determined by the DBI action. However, in this paper, we will
simply consider the D7 brane infinitely far away from the horizon, which corresponds to an infinitely heavy quark. 

For a pair of parallel strings, the embedding is simply
\begin{equation}
  x(r) = \mathrm{constant} \,, 
\end{equation}
and for a U-shaped strings, we have the conserved quantity ($x$ is a cyclic coordinate) 
\begin{equation} \label{cyclic1}
  c_0 \equiv \frac{e^{\phi(r)/2}}{\sqrt{\frac{r^4}{R^4}f(r) + 1/x'^2}} \frac{r^4}{R^4} f(r) = \frac{e^{\phi(r_0)/2} r_0^2 \sqrt{f(r_0)}}{R^2} \,,
\end{equation}
where $r_0$ is the minimum value of $r$ giving $x'(r_0) = \infty$. 
From  \eqref{cyclic1} we have
\begin{equation} \label{x1}
  x(r) = 2 R^2 \int^{r}_{r_0} \dd r \frac{1}{r^2 \sqrt{f(r)}
  \sqrt{e^{\phi(r)}r^4 f/(e^{\phi(r_0)}r_0^4f(r_0))   -1}} 
\end{equation}
We can compute the on-shell action of the U-shaped string embedding ($S_{NG}^{\mathrm{U}}$) by plugging \eqref{x1} into \eqref{NG1}.
However, it is divergent because of the infinitely heavy quark mass contribution. We regularize the on-shell action by subtracting this infinite mass, which is nothing but the on-shell action of a pair of parallel strings ($S_{NG}^{\parallel}$). 

\begin{figure}[]
\centering
   \subfigure[$A=0$, $w_H=0$ (dotted line) and $w_H=0.5$ (solid line)]
   {\includegraphics[width=6cm]{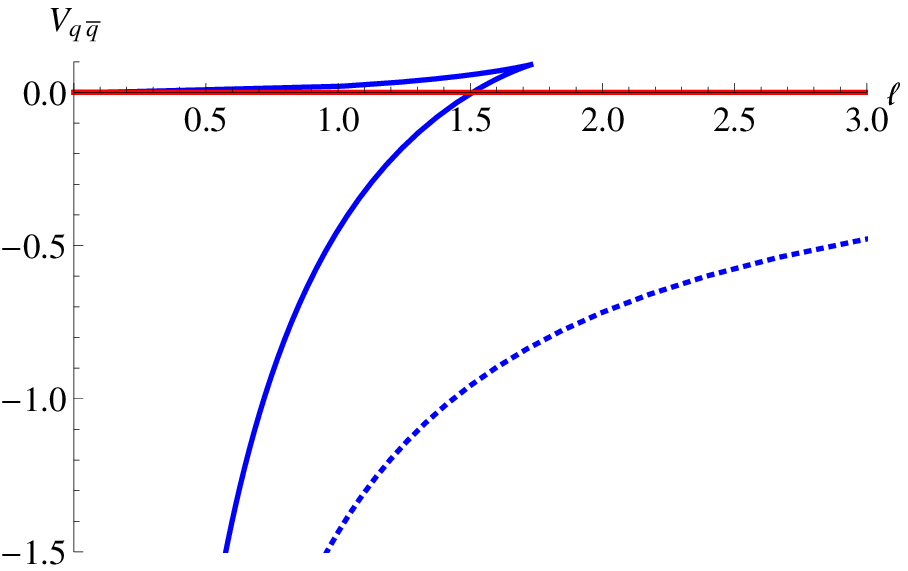} }
    \subfigure[$w_H=0.5$, $A=0,1,3,5,10$ from top ]
   {\includegraphics[width=6cm]{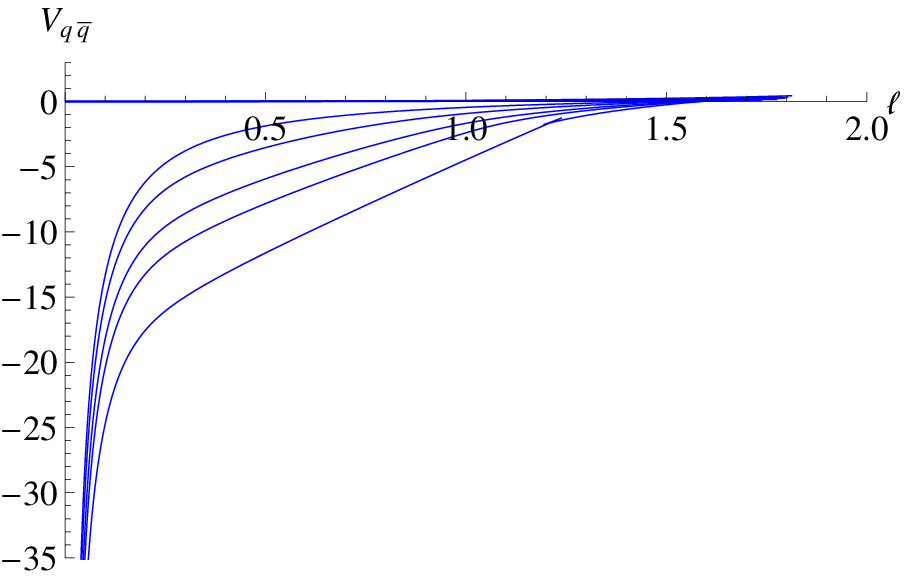} }
     \subfigure[$A=3$, $w_H=2,1,0.5,0.2$ from top ]
   {\includegraphics[width=6cm]{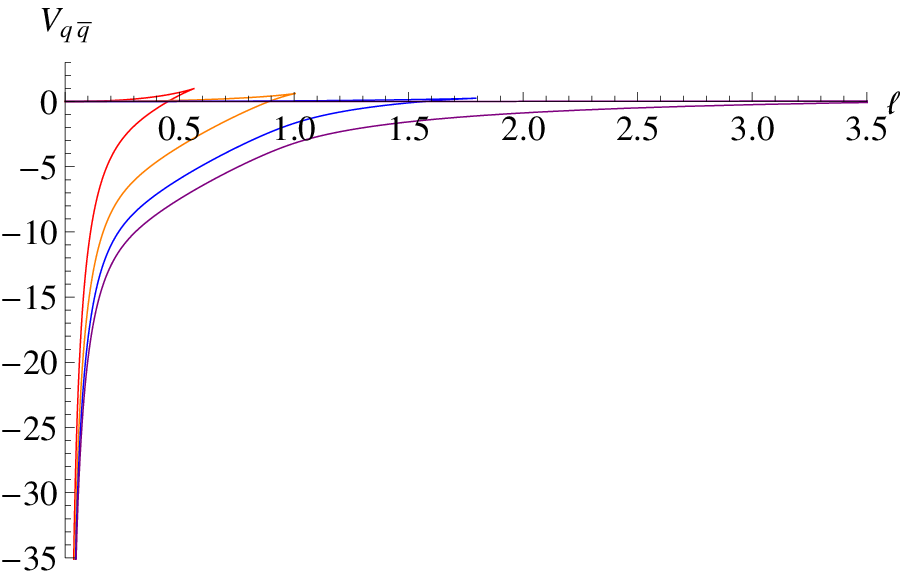} }
    \subfigure[$A=10$, $w_H=2,1,0.5,0.2$ from top ]
   {\includegraphics[width=6cm]{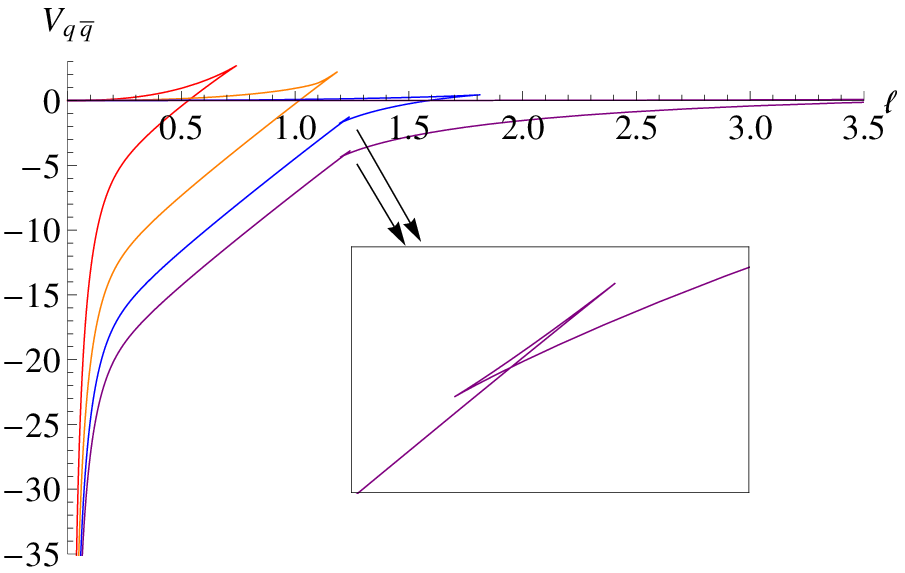}} 
  \caption{Quark-antiquark potential ($\Gamma=1, \lambda = 1.715$)} \label{wilson}
\end{figure}

In summary, we define the quark-antiquark potential ($V_{q\bar{q}}(l)$) as 
\begin{equation}
\begin{split}
  V_{q\bar{q}}(l) &\equiv \frac{S_{NG}^{\mathrm{U}} - S_{NG}^{\parallel}}{2\pi\a'} \\
   &= 2 \int_{r_0}^\infty \dd r \frac{e^{\phi(r)/2}}{\sqrt{1- e^{\phi(r_0)} r_0^4 f(r_0) / (e^\phi(r) r^4 f(r)) }} 
  - 2 \int_{r_H}^\infty \dd r  e^{\phi(r)/2} \,.
  \end{split}
\end{equation}
which is the function of $l \equiv x(\infty)$, the distance between the quark and antiquark. $r_0$ on the right hand side can be related to 
$l = x(\infty) $ by \eqref{x1}. 
Our numerical plots of $V_{q\bar{q}}$ for given parameter sets are shown in Fig \ref{wilson}.

Let us start with the case $A=0$ which is just the well known ${\cal N}=4$ theory, Fig \ref{wilson}(a). 
This is the case with 
a trivial dilaton, $e^\phi=1$, and studied first in \cite{Rey:1998ik,Maldacena:1998im,Rey:1998bq,Brandhuber:1998bs}. At zero $T$, the 
potential scales as $1/l$ (dotted line), which is a consequence of the conformality. 
At finite low $T$, the potential scales as $1/l(1+ \calo((Tl)^4))$. Again 
the dependence on the $Tl$ combination is due to the underlying conformal symmetry. The new feature at finite $T$ is the existence of the phase transition from the bound quark antiquark pair to the free quark state, as the distance between quarks increases. In Fig \ref{wilson}(a) it corresponds roughly to the transition from the blue solid curve (a U-shape string) to the red horizontal line (a pair of parallel strings)  near $l=1.5$\footnote{
In principle one should also take into 
account  ``graviton exchange'' between the separated string worldsheets as was done carefully in \cite{Bak:2007fk} for  large quark separation. This will modify the potential at large separation and will
replace the phase transition shown at $l \sim 1.5$ with a cross-over.
Since we focus on the dilaton effect at intermediate distance 
we don't consider that modification. If we considered it 
we could have a smoother cross over at larger distance, and also possibly, 
lose the additional first order transition in the inset of Fig \ref{wilson}(d).
}.

Now let us turn on A at a fixed T, Fig 7(b) ($w_H=0.5$ and here again we set the scale  in \eqref{dila1} $\lambda=1.715$).
As $A$ increases a linear potential starts forming, which is the 
characteristic feature of confinement. 
It is interesting that the linearity is very clear at $A=10$, where 
there exists a stable baryon phase Fig \ref{final}(b)\footnote{This linear potential was observed also in a similar setup with $e^\phi = 1- q/r_H^4 \log\left(1-r_H^4/r^4\right)$, where $q$ plays a role of our $A$ \cite{Ghoroku:2005tf}.}. So, we see a rough correlation 
between our phase diagrams Fig \ref{final} and $V_{q\bar{q}}$ (Fig \ref{wilson}). 
Fig \ref{wilson}(c) corresponds to Fig \ref{final}(a). The linearity 
is not clear in Fig \ref{wilson}(c) and it is natural that we expect a deconfined phase and no stable baryons as shown in Fig \ref{final}(a). Fig \ref{wilson}(d) corresponds to Fig \ref{final}(b). At high temperature we tend to loose the linearity and there is no big difference from Fig \ref{wilson}(c) (deconfinment, no baryon).
However at lower temperature, the linearity becomes stronger (confinement, baryon phase). So it is consistent with the phase diagram Fig \ref{final}(b).   

There is one additional interesting feature at low temperature and large $A$. 
As shown in the insert of Fig \ref{wilson}(d), there is a first order phase
transition between the linear potential and Coulomb-like potential.
This is at odds with QCD, but makes good sense in our model.  
Our dilaton profile is a step function so there is a potential barrier localized at $r=\l$. As $l$ increases, the string will extend  
inside, reduce $r_0$ and finally meet $r_0 = \l$, where a linear potential 
is built. If we further increase $l$, then the string finally will 
manage to get through the barrier at $r=\l$ and beyond that point, 
there is no friction against the string moving to lower $r$ since the dilaton is constant. 
Therefore the potential becomes Coulomb-like again. 
We don't see this first order transition at high $T$, because at high $T$, 
the interval between $r_H$ and $\l$ is too short and there is no room 
for a ``Coulomb'' phase. i.e. After getting inside $r=\l$, the string should meet the black hole horizon very quickly. 

The Coulomb-phase should not be present in QCD. 
We could easily cure this artifact by considering a continuously growing dilaton profile in the IR rather than the flat
form. However, to keep the phase diagram seen in Fig \ref{final}(b), the 
dilaton should only increase mildly (For example, as $1/r^q$ as studied in \cite{Alvares:2012kr}). 
If it is too strong, the phase diagram will be always the type of Fig \ref{final}(c), destroying 
our motivation for this paper. However, a simpler explanation of the Coulomb phase is that it is a result of
the quenched approximation. In reality we would expect $\bar{q}q$ pair production to break the string at
separations before the Coulomb phase sets in for large $A$. For this reason these models still seem reasonable for QCD.
In this section, our discussion has only been qualitative. More 
quantitative studies of the Wilson-Polyakov loop, together with other modifications of dilaton, would be interesting.

\section{Summary}

In this paper we have used the D3/D7 system to holographically study the phase diagram of a chiral symmetry
breaking gauge theory as a function of the running coupling profile. We have included a running coupling
through a phenomenological non-backreacted dilaton profile which steps between conformal UV and IR regimes. 
Here we have considered dependence on the height of the step. The model has previously been shown to
have three phases \cite{Evans:2011eu}: a chirally symmetric quark plasma at high T and $\mu$; a chiral symmetry broken phase
at small $T$ and $\mu$; and a more exotic chiral symmetry broken phase with quark density at intermediate
$T,\mu$ for some parameter values. The order of these phase transitions depend on the height of the step.
A first order transition to the chirally symmetric phase can be achieved for low step values. Previous
work has also shown that the low $\mu$ transition with $T$ can be made second order by phenomenologically
tinkering with the shape of the AdS black hole horizon (in a way compatible with the spatial symmetry breaking of
the D3/D7 system). 

The crucial extra ingredient we have concentrated in this analysis is the low $T$ and intermediate $\mu$ baryonic 
phase. Baryons can be introduced as D5 branes wrapped on the $S^5$ of the dual geometry and linked D7-D5
systems describe the hadronic phase. In \cite{Gwak:2012ht} such configurations were introduced in a  geometry
with a diverging dilaton in the IR. The resulting baryonic phase persisted though to arbitrarily high density
unlike in QCD (but in a similar fashion to the equivalent phase diagram in the Sakai Sugimoto model 
\cite{Bergman:2007wp}). Our intuition for our analysis was that in a model with a step function dilaton profile there
would be no very large or small baryon vertices because in the two conformal regimes they would shrink away. The
only D5 brane configurations would lie around the step and we could hope they would only play a role at intermediate
$\mu$. We indeed find, after careful numerical analysis, that this is the case for some range of the step size.
Our model which best matches expectations for QCD in the low $T$ regime is shown in Fig \ref{final}(b). The
second order transition to this  regime matches expectations in the absence of internuclear interactions
which we neglected. 

We have also made a qualitative link between the presence of the baryonic phase and  
the observation of linear confinement in Wilson loop computations in the background. The step function dilaton
form generates both confinement and chiral symmetry breaking when the step size is large enough.

The phase diagrams we produce are not intended as predictions for QCD since the underlying physics model is
somewhat different (eg the presence of super-partners) but they do demonstrate the wealth of behaviours
possible in strongly coupled gauge theory. We can also hope for some universality and by finding models
that match QCD's expectations for the phase structure it may be possible in the future 
to study phenomena beyond that stucture using the models. For example in \cite{Evans:2010xs} the temporal behaviour of 
these systems, such as bubble formation,
could be followed through a first order phase transition in a model of this type.

\acknowledgments
NE is grateful for the support of
an STFC rolling grant. MM is grateful for University of Southampton Scholarships. KK acknowledge support via an NWO Vici grant of K. Skenderis  and University of Southampton Scholarships. This work is part of the research program of the Stichting voor Fundamenteel Onderzoek der Materie (FOM), which is financially supported by the Nederlandse Organisatie voor Wetenschappelijk Onderzoek (NWO). The work of YS and SJS was supported by Mid-career Researcher Program through NRF grant(No. 2010-0008456).



\providecommand{\href}[2]{#2}\begingroup\raggedright\endgroup

\end{document}